\documentclass[preprint,prc,showpacs,preprintnumbers,
               superscriptaddress,amsmath,amssymb,floatfix]{revtex4}

\usepackage{graphicx,epsfig,latexsym,amssymb}
\usepackage{multirow,amsmath,array,booktabs}
\usepackage{dcolumn}
\usepackage[section]{placeins}
\usepackage{bm}

\newcommand{\ls}{\left[}
\newcommand{\rs}{\right]}
\newcommand{\pade}{\mbox{Pad$\acute{\text{e}}$}}
\newcommand{\svec}[1]{{\mbox{\boldmath${\rm #1}$}}}

\newcommand{\re}{\nonumber\\}

\renewcommand{\emph}[1]{\textit{\textbf{#1}}}

\begin{document}

\title{Analytic continuation of single-particle resonance energy and  wave function in relativistic
       mean field theory}

\author{S. S. Zhang}
 \affiliation{School of Physics, Peking University, Beijing 100871,
              China}

\author{J. Meng}
 \email{mengj@pku.edu.cn}
 \affiliation{School of Physics, Peking University, Beijing 100871,
              China}
 \affiliation{Institute of Theoretical Physics, Chinese Academy of
              Sciences, Beijing 100080, China}
 \affiliation{Center of Theoretical Nuclear Physics, National Laboratory
              of Heavy Ion Accelerator, Lanzhou 730000, China}

 \author{S. G. Zhou}
 \affiliation{School of Physics, Peking University, Beijing 100871,
              China}
 \affiliation{Institute of Theoretical Physics, Chinese Academy of
              Sciences, Beijing 100080, China}
 \affiliation{Center of Theoretical Nuclear Physics, National Laboratory
              of Heavy Ion Accelerator, Lanzhou 730000, China}

 \author{G. C. Hillhouse}

\affiliation{Department of Physics, University
of Stellenbosch, Private Bag X1, Matieland 7602, South Africa}
 \affiliation{School of Physics, Peking University, Beijing 100871,
              China}

\begin{abstract}

Single-particle resonant states in spherical nuclei are studied by
an analytic continuation in the coupling constant (ACCC) method
within the framework of the self-consistent relativistic mean
field (RMF) theory. Taking the neutron resonant state $\nu
1g_{9/2}$ in $^{60}$Ca as an example, we examine the analyticity
of the eigenvalue and eigenfunction for the Dirac equation with
respect to the coupling constant by means of a $\pade$ approximant
of the second kind. The RMF-ACCC approach is then applied to
$^{122}$Zr and, for the first time, this approach is employed to
investigate both the energies, widths and wave functions for $l\ne
0$ resonant states close to the continuum threshold. Predictions
are also compared with corresponding results obtained from the
scattering phase shift method.

\end{abstract}
 \maketitle

\section{Introduction}

The physics of exotic nuclei with unusual $N/Z$ ratios (isospin)
has attracted world-wide attention. These nuclei are usually
loosely bound and often exhibit resonances with a pronounced
single-particle character since the Fermi surface approaches the
continuum. In some nuclei, the valence nucleons can be easily
scattered into single-particle resonant states (or Gamow states)
in the continuum which is essential for the existence of exotic
phenomena. For example, the halo in $^{11}$Li~\cite{TA85} can be
successfully reproduced~\cite{MR96} and giant halo
nuclei~\cite{MR98} are predicted by the relativistic continuum
Hartree-Bogoliubov (RCHB) theory~\cite{ME98} in which the
contribution from the continuum has been proven to be crucial to
understand these halo phenomena. There are also other attempts to
include the contribution from the continuum in the
non-relativistic~\cite{San00} and relativistic
approaches~\cite{San}. By taking into account a few resonant
states close to the continuum threshold with the BCS
method~\cite{San}, the relativistic mean field (RMF) theory seems
to be able to reproduce the RCHB results and describe the main
effects of the continuum. It has also been shown that the
contribution of the continuum to the giant resonances mainly comes
from Gamow states~\cite{Curutchet89,Cao02}. Therefore, a proper
treatment of resonant states is important for a deeper
understanding of the properties in exotic nuclei.

Presently, several techniques have been developed to study
resonant states in the continuum. One of them is the R-matrix
theory in which resonance parameters (i.e. energy and width) can
be reasonably determined from fitting the available experimental
data~\cite{Wigner47}. The extended R-matrix theory~\cite{Hale87}
and the K-matrix theory~\cite{Humblet91} have also been developed.
The conventional scattering theory is also an efficient tool for
studying resonances. More precisely, the phase shift method is
commonly used to determine parameters for a resonant state which
corresponds to a pole of the S matrix~\cite{Taylor72}. By
discretizing the continuum, the contribution of the resonant
states can be self-consistently taken into account via a
Bogoliubov transformation in coordinate
space~\cite{Dobaczewski84,MR96}.

Computationally, it is desired to deduce the properties of unbound
states from the eigenvalues and eigenfunctions of Hamiltonians for
bound states so that the methods developed for bound states can
still be used. For this purpose, the bound-state-type methods have
been developed, including the real stabilization
method~\cite{Hazi70}, the complex scaling method
(CSM)~\cite{Ho83}, and the analytic continuation in the coupling
constant (ACCC) method~\cite{kukulin89}. In the real stabilization
method, the continuum is discretized in a box and the resonance
can be found according to the fact that its energy should be
stable against the change of the box size. Sometimes it is
difficult to find the appropriate box sizes, and it also turns out
that this approach is suitable for narrow resonances
only~\cite{tanaka99}. In the CSM, the wave function for a resonant
state is transformed to the square-integrable wave function for a
bound state by rotating the position vector in coordinate space
into the complex plane. Since the CSM involves the calculations of
complex matrix elements, the corresponding difficulties related to
complex eigenvalue problem still exist. In the ACCC method, a
resonant state becomes a bound state if one increases the
attractive potential and the energy, width and wave function for
the resonant state can be obtained by an analytic continuation
carried out via a $\pade$ approximant (PA) from the bound-state
solutions. Compared with other bound-state-type methods, the ACCC
approach is very effective and numerically quite simple because
many methods available for bound-state problems can be used and
the PA for analytic continuation can be implemented easily.

Within the framework of the non-relativistic Schr\"{o}dinger
equation, the ACCC method has been applied to investigate the
energies and widths for resonant states in light nuclei combined
with few-body methods~\cite{tanaka99,Aoyama02}, and to study
single-particle resonant states in spherical and deformed nuclei
by solving the Schr\"{o}dinger equation with Woods-Saxon
potentials~\cite{cattapan}. Based on the Schr\"{o}dinger and Dirac
equations, the stability and convergence of the energies and
widths for single-particle resonant states with square-well,
harmonic-oscillator and Woods-Saxon potentials have been
investigated and their dependence on the coupling constant
interval and the order of the PA have been discussed in
detail~\cite{ZhangSch,ZhangDirac}.

The relativistic mean field (RMF) theory is widely used to
describe properties of nuclear matter and finite nuclei
successfully, not only for those nuclei near the valley of
stability, but also for exotic nuclei with large neutron (proton)
excess~\cite{Ring96,MR96}. Previously the ACCC method was employed
to study energies and widths for resonant states in stable nuclei
$^{16}$O and $^{48}$Ca in RMF theory~\cite{Yang01}. In the present
work, we aim at exploring the single-particle resonant states,
including resonance parameters and wave functions, in exotic
nuclei by the ACCC method within the framework of the RMF theory.
This paper represents the first application of the RMF-ACCC method
to obtain both the wave functions and resonance parameters. The
paper is organized as follows. In Sec. II the ACCC method combined
with the RMF theory is presented. The numerical details are given
in Sec. III and its application to the doubly magic nucleus
$^{122}$Zr in Sec. IV. Finally, we give a brief summary in Sec. V.

\section{Theoretical framework}

\subsection{Relativistic Mean Field Theory}

The basic ansatz of the RMF theory is a Lagrangian density whereby
nucleons are described as Dirac particles which interact via the
exchange of various mesons (the scalar $\sigma$, vector $\omega$
and iso-vector vector $\rho$) and the photon~\cite{ME98}
\begin{eqnarray}
\begin{array}{cc}
{\cal L} &= \bar \psi (i\rlap{/}\partial -M) \psi +
        \,{1\over2}\partial_\mu\sigma\partial^\mu\sigma-U(\sigma)
        -{1\over4}\Omega_{\mu\nu}\Omega^{\mu\nu} \\
        \ &+ {1\over2}m_\omega^2\omega_\mu\omega^\mu
        -{1\over4}{\svec R}_{\mu\nu}{\svec R}^{\mu\nu} +
        {1\over2}m_{\rho}^{2} \svec\rho_\mu\svec\rho^\mu
        -{1\over4}F_{\mu\nu}F^{\mu\nu} \\
        & -  g_{\sigma}\bar\psi \sigma \psi~
        -~g_{\omega}\bar\psi \rlap{/}\omega \psi~
        -~g_{\rho}  \bar\psi
        \rlap{/}\svec\rho
        \svec\tau \psi
        -~e \bar\psi \rlap{/}\svec A \psi,
\label{Lagrangian}
\end{array}
\end{eqnarray}
where $M$ is the nucleon mass and $m_\sigma$ ($g_\sigma$),
$m_\omega$ ($g_\omega$), and $m_\rho$ ($g_\rho$) are the masses
(coupling constants) of the respective mesons. A nonlinear scalar
self-interaction $U(\sigma)~=~\dfrac{1}{2} m^2_\sigma \sigma^2
~+~\dfrac{g_2}{3}\sigma^3~+~\dfrac{g_3}{4}\sigma^4$ of the
$\sigma$ meson has been included~\cite{BB77}. The field tensors
for the vector mesons are given as
\begin{eqnarray}
\left\{
\begin{array}{lll}
   \Omega^{\mu\nu}   &=& \partial^\mu\omega^\nu-\partial^\nu\omega^\mu, \\
   {\svec R}^{\mu\nu} &=& \partial^\mu{\svec \rho}^\nu
                        -\partial^\nu{\svec \rho}^\mu
                        - g^{\rho} ( {\svec \rho}^\mu
                           \times {\svec \rho}^\nu ), \\
   F^{\mu\nu}        &=& \partial^\mu \svec A^\nu-\partial^\nu \svec A^\mu.
\end{array}   \right.
\label{tensors}
\end{eqnarray}

The classical variation principle gives the following equations of
motion
\begin{equation}
   [ {\svec \alpha} \cdot {\svec p} +
     V_V ( {\svec r} ) + \beta ( M + V_S ( {\svec r} ) ) ]
     \psi_i ~=~ \epsilon_i\psi_i,
\label{spinor1} \end{equation} for the nucleon spinors, where
$\varepsilon_i$ and $\psi_i$ are the single-particle energy and
spinor wave function respectively, and the Klein-Gordon equations
\begin{eqnarray}
\left\{
\begin{array}{lll}
  \left( -\Delta \sigma ~+~U'(\sigma) \right ) &=& g_\sigma\rho_s,
\\
   \left( -\Delta~+~m_\omega^2\right )\omega^{\mu} &=&
                    g_\omega j^{\mu} ( {\svec r} ),
\\
   \left( -\Delta~+~m_\rho^2\right) {\svec \rho}^{\mu}&=&
                    g_\rho \svec j^{\mu}( {\svec r} ),
\\
          -\Delta~ A_0^{\mu} ( {\svec r} ) ~ &=&
                           e j_{\rho}^{\mu}( {\svec r} ),
\end{array}  \right.
\label{mesonmotion}
\end{eqnarray}
for the mesons, where
\begin{eqnarray}
\left\{
\begin{array}{lll}
   V_V( {\svec r} ) &=&
      g_\omega\rlap{/}\omega + g_\rho\rlap{/}\svec\rho\svec\tau
         + \dfrac{1}{2}e(1-\tau_3)\rlap{\,/}\svec A , \\
   V_S( {\svec r} ) &=&
      g_\sigma \sigma , \\
\end{array}
\right. \label{vaspot}
\end{eqnarray}
are the vector and scalar potentials respectively and the source
terms for the mesons are
\begin{eqnarray}
\left\{
\begin{array}{lll}
   \rho_s &=& \sum\limits_{i=1}^A \bar\psi_i \psi_i,
\\
   j^{\mu} ( {\svec r} ) &=&
               \sum\limits_{i=1}^A \bar \psi_i \gamma^{\mu} \psi_i,
\\
   \svec j^{\mu}( {\svec r} ) &=&
          \sum\limits_{i=1}^A \bar \psi_i \gamma^{\mu} \svec \tau
          \psi_i,
\\
   j^{\mu}_p ( {\svec r} ) &=&
      \sum\limits_{i=1}^A \bar \psi_i \gamma^{\mu} \dfrac {1 - \tau_3} 2
      \psi_i .
\end{array}  \right.
\label{mesonsource}
\end{eqnarray}
It should be noted that the contribution of negative energy states
are neglected, i.e., the vacuum is not polarized. Moreover, the
mean field approximation is carried out via replacing meson field
operators in Eq. (\ref{spinor1}) by their expectation values,
since the coupled equations Eq. (\ref{spinor1}) and Eq.
(\ref{mesonmotion}) are nonlinear quantum field equations and
their exact solutions are very complicated. In this way, the
nucleons are assumed to move independently in the classical meson
fields. The coupled equations can be solved self-consistently by
iteration.
\par

For spherical nuclei, the potential of the nucleon and the sources
of meson fields depend only on the radial coordinate $r$. The
spinor is characterized by the angular momentum quantum numbers
$l$, $j$, $m$, the isospin $t = \pm \dfrac 1 2$ for neutron and
proton respectively, and the other quantum number $i$. The Dirac
spinor has the form
\begin{equation}
   \psi ( \svec r ) =
      \left( { {\mbox{i}  \dfrac {G_i^{lj}(r)} r {Y^l _{jm} (\theta,\phi)} }
      \atop
       { \dfrac {F_i^{lj}(r)} r (\svec\sigma \cdot \hat {\svec r} )
       {Y^l _{jm} (\theta,\phi)} } }
      \right) \chi _{t}(t),
\label{reppsi}
\end{equation}
where $Y^l _{jm} (\theta,\phi)$ are the spinor spherical
harmonics. The radial equation of the spinor, i.e. Eq.
(\ref{spinor1}), can be reduced to ~\cite{ME98}
\begin{eqnarray}
  \label{Dirac-r}
         (-\dfrac{\partial}{\partial
          r}+\dfrac{\kappa}{r})F_i^{lj}(r)+\ls M+V_p(r)\rs
          G_i^{lj}(r)&=&\varepsilon_i G_i^{lj}(r),\re
         (\dfrac{\partial}{\partial r}+\dfrac{\kappa}{r})G_i^{lj}(r)-\ls
         M-V_m(r)\rs F_i^{lj}(r)&=&\varepsilon_i F_i^{lj}(r),
\end{eqnarray}
in which $V_p(r)=V_V(r)+V_S(r)$, $V_m(r)=V_V(r)-V_S(r)$, and
$\kappa=(-1)^{j+l+1/2}(j+1/2)$. The meson field equations can be
reduced to
\begin{equation}
    \left( \frac {\partial^2} {\partial r^2}  - \frac 2 r \frac
         {\partial}   {\partial r} + m_{\phi}^2 \right)\phi = s_{\phi} (r),
    \label{Ramesonmotion}
\end{equation}
where $\phi = \sigma$, $\omega$, $\rho$, and photon ( $m_{\phi} =
0$ for photon). The source terms read
\begin{eqnarray}
  s_{\phi} (r) = \left\{
    \begin{array}{ll}
      -g_\sigma\rho_s - g_2 \sigma^2(r)  - g_3 \sigma^3(r)
           & { \rm for ~ the ~  \sigma~  field }, \\
      g_\omega \rho_v    & {\rm for ~ the ~ \omega ~ field}, \\
      g_{\rho}  \rho_3(r)       & {\rm for~ the~ \rho~ field},\\
      e \rho_c(r)  & {\rm for~ the~ Coulomb~ field}, \\
   \end{array} \right.
\end{eqnarray}
and
\begin{eqnarray}
\left\{
\begin{array}{lll}
   4\pi r^2 \rho_s (r) &=& \sum\limits_{i=1}^A ( |G_i(r)|^2 - |F_i(r)|^2 ), \\
   4\pi r^2 \rho_v (r) &=& \sum\limits_{i=1}^A ( |G_i(r)|^2 + |F_i(r)|^2 ), \\
   4\pi r^2 \rho_3 (r) &=& \sum\limits_{p=1}^Z ( |G_p(r)|^2 + |F_p(r)|^2
   )
               -  \sum\limits_{n=1}^N ( |G_n(r)|^2 + |F_n(r)|^2 ), \\
   4\pi r^2 \rho_c (r) &=& \sum\limits_{p=1}^Z ( |G_p(r)|^2 + |F_p(r)|^2 ) . \\
\end{array}
\right.
\label{mesonsourceS}
\end{eqnarray}

\subsection{Analytic Continuation
in the Coupling Constant for Resonance Parameters and Gamow Wave
Functions}

By solving Eqs.~(\ref{Dirac-r}) and (\ref{Ramesonmotion}) in a
meshed box of size $R_0$ self-consistently, one can calculate the
ground state properties of a nucleus. The vector potential
$V_V(r)$ and the scalar potential $V_S(r)$, energies and wave
functions for bound states are also obtained. By increasing the
attractive potential as $V_p(r)\rightarrow\lambda V_p(r)$, a
resonant state will be lowered and becomes a bound state if the
coupling constant $\lambda$ is large enough. Near the branch point
$\lambda_0$, defined by the scattering threshold
$k(\lambda_0)=0$~\cite{kukulin89}, the wave number $k(\lambda)$
behaves as
\begin{eqnarray}
   k(\lambda) \sim \left\{
   \begin{array}{l@{\mbox{~~~}}l}
         i\sqrt{\lambda-\lambda_0},        &  l>0 ,\\
         i(\lambda-\lambda_0),            &  l=0.
   \end{array}\right.
\end{eqnarray}
These properties suggest an analytic continuation of the wave
number $k$ in the complex $\lambda$ plane from the bound-state
region into the resonance region by $\pade$ approximant of the
second kind (PAII)~\cite{kukulin89}
\begin{equation}
    \label{pade-e}
    k(x)\approx k^{[L,N]}(x)= i \dfrac{c_0+c_1 x+c_2x^2+\ldots +c_Lx^L}{ 1+d_1 x+d_2 x^2+\ldots+d_N x^N},
\end{equation}
where $x\equiv\sqrt{\lambda-\lambda_0}$, and $c_0, c_1,\ldots,
c_L, d_1, d_2,\ldots,d_N$ are the coefficients of PA. These
coefficients can be determined by a set of reference points $x_i$
and $k(x_i)$ obtained from the Dirac equation with
$\lambda_i>\lambda_0,~i=1,2,...,L+N+1$. With the complex wave
number $k(\lambda=1)= k_r + i k_i$, the resonance energy $E$ and
the width $\Gamma$ can be extracted from the relation $
\varepsilon=E-i \dfrac{\Gamma}{2} ~~ (E,\Gamma\in \mathbb{R})$ and
$k^2=\varepsilon^2-M^2$, i.e.,
\begin{eqnarray}
    \label{E-W}
    E      &=& \sqrt{\dfrac{\sqrt{(M^2+k_r^2-k_i^2)^2+4k_r^2k_i^2}+(M^2+k_r^2-k_i^2)}{2}}- M , \re
    \Gamma &=& \sqrt{2\sqrt{(M^2+k_r^2-k_i^2)^2+4k_r^2k_i^2}-2(M^2+k_r^2-k_i^2)}.
\end{eqnarray}
In the non-relativistic limit ($k\ll M$), Eq. (\ref{E-W}) reduces
to
\begin{equation}
    E=\dfrac{k_r^2-k_i^2}{2M},~~~~~~~~~~~~~~~~~
    \Gamma=\dfrac{2k_rk_i}{M}.
\end{equation}

It is evident that the continuation in the coupling constant can
be replaced by the continuation in $k$ plane along the
$k(\lambda)$ trajectory determined by Eq. (\ref{pade-e}) to the
point $k_R$ corresponding to the wave number for Gamow state,
i.e., $k_R=k^{[L,N]}(\lambda=1)$~\cite{kukulin89}. Similarly, the
wave function $\varphi(k_R,r)$ for a resonant state can be
obtained by an analytic continuation of the bound-state wave
function $\varphi(k_i,r)$ in the complex $k$ plane. One can also
prove that the wave function $\varphi(k,r)$ is an analytic
function of the wave number $k$ in the inner region $r<R_0$ where
the Jost function analyticity dominates~\cite{kukulin89}.
Therefore, we use the technique, which has been adopted to find
the complex resonance energy, to determine the resonance wave
function $\varphi(k_R,r)$. Firstly, we construct the PA to define
the resonance wave function at any point $r$ in the inner region
($r<R_0$)~\cite{kukulin89}
\begin{equation}
    \label{inner}
    \varphi^{[L,N]}(k,r) = \dfrac{P_L(k,r)}{Q_N(k,r)}
                         =\dfrac{a_0(r)+a_1(r)k+a_2(r)k^2+\ldots +a_L(r)k^L}{ 1+b_1(r) k+b_2(r)k^2+\ldots+b_N(r) k^N},
\end{equation}
where the coefficients ${a_i(r)}~(i=0,1,\ldots, L)$ and
${b_j(r)}~(j=1,2,\ldots, N)$ are dependent on $r$. These
coefficients can be determined by a set of reference points $k_i$
and $\varphi(k_i,r)$ obtained from the Dirac equation with
$\lambda_i>\lambda_0,~(i=1,2,...,L+N+1)$. The resonance wave
function $\varphi(k_R,r)=\varphi^{[L,N]}(k_R,r)~(r<R_0)$ can be
extrapolated in this way.

Eq. (\ref{Dirac-r}) can be rewritten as two decoupled
Schr\"odinger-like equations for the upper and the lower
components respectively~\cite{ME99}. For neutrons, the
Schr\"odinger-like equation for the upper component reads
\begin{equation}
  \label{G-diff}
  \dfrac{\partial^2G_i^{lj}(r)}{\partial
  r^2}-\dfrac{\kappa_i(\kappa_i+1)}{r^2}G_i^{lj}(r)+(\varepsilon_i^2-M^2)G_i^{lj}(r)=0,
\end{equation}
in the outer region where $V_V(r)\simeq 0$ and $V_S(r)\simeq 0$.
Its solution is the well known Riccati-Hankel function
\begin{equation}
    \hat{h}^{\pm}_\kappa (z)=\hat{n}_\kappa (z)\pm i\hat{j}_\kappa(z), ~~~~~z=kr,
\end{equation}
where $ \hat{j}_\kappa(z)=z{j}_\kappa (z)$ is the usual regular
solution, i.e. Riccati-Bessel function, and
$\hat{n}_\kappa(z)=z{n}_\kappa (z)$ is the irregular solution,
i.e. Riccati-Neumann function. The outer wave function is matched
to the inner wave function at $r = r_m < R_0$
\begin{equation}
   \label{match-G}
   \varphi^{[L,N]}(k_R,r_m) = C(k_R)\ls\hat{j}_\kappa(k_Rr_m)+ D(k_R)\hat{n}_\kappa(k_Rr_m)\rs,
\end{equation}
where $C(k_R)$ is the coefficient for matching and
$D(k_R)=\tan\delta_\kappa(k_R)$ with $\delta_\kappa(k_R)$ the
phase shift. It has been found that $\delta_\kappa(k_R)$ is almost
a constant when $r_m$ is large enough. Given the upper component,
the lower component $F_{\kappa}(r)$ can be calculated from the
relationship between the upper and the lower components derived
from Eq.~(\ref{Dirac-r}), i.e.,
\begin{equation}
   \label{match-F}
   F_{\kappa}(r) = \dfrac{(\varepsilon-m)}{k_RC(k_R)}\ls
   \hat{j}_{\kappa-1}(k_Rr)+D(k_R)\hat{n}_{\kappa-1}(k_Rr)\rs.
\end{equation}
Finally, the resonance wave function is normalized according to
the Zel'dovich procedures~\cite{kukulin89}.

\section{Numerical details}

In this section we give the details on how the ACCC method is
realized within the framework of the RMF theory and exploit the
analyticity of the eigenvalues and eigenfunctions of the Dirac
equation with respect to the coupling constant. In RCHB
calculations $^{60}$Ca is the doorway for exotic phenomena such as
giant halos~\cite{MR98}, in which the neutron resonant states play
an important role in the continuum. We select the neutron resonant
state $\nu 1g_{9/2}$ in $^{60}$Ca to investigate the analyticity
of the eigenvalue and eigenfunction for the Dirac equation with
respect to the coupling constant. The RMF equations are solved in
a spherical box of the size $R_0 = 20$ fm, with a mesh size of
$\delta r = 0.05$ fm~\cite{ZMR03}. Several well used effective
interactions for the RMF Lagrangian will be used and the
corresponding results will be compared. The energy, width and wave
function for the $\nu 1g_{9/2}$ resonant state in $^{60}$Ca are
obtained with the procedure given in the last section. As the
stability and convergence (e.g., the dependence on PA order) of
the energies and widths in the Dirac and Schr\"{o}dinger equations
have been investigated in Refs.~\cite{ZhangSch,ZhangDirac}, we do
not discuss them here and use $L=N=4$ PA for the present
calculations.

In Tab.~\ref{tab1}, we present the energies and widths for $\nu
1g_{9/2}$ in $^{60}$Ca calculated with effective interactions
NL3~\cite{NL3}, NLSH~\cite{NLSH}, TMA~\cite{TMA} and the newly
developed PK1~\cite{PK1} parametrization. The results from the
scattering phase shift method (RMF-S) obtained similarly as in
Ref.~\cite{San} are also included for comparison. The first step
in  RMF-S is to solve the Dirac equation [Eq. (\ref{Dirac-r})]
with scattering boundary conditions for the continuum spectrum.
The solutions, upper and lower components, have the same form as
here in Eq. (\ref{match-G}) and Eq. (\ref{match-F}) respectively.
The coefficients $C$ and $D$  in Eq. (\ref{match-G}) and Eq.
(\ref{match-F}) are fixed by the normalization condition of the
scattering wave functions and the phase shift $\delta$ is
calculated from the matching conditions. It should be mentioned
that the energy used in the wave functions is real and not unique,
as opposed to complex and the unique value which is determined by
the ACCC method in advance. In the vicinity of an isolated
resonance, the resonance energy and width are extracted from the
derivative of the phase shift $\delta$ in Breit-Wigner form, i.e.,
\begin{equation}
    \dfrac{d\delta(E)}{dE}=\dfrac{\Gamma/2}{(E_r-E)^2+\Gamma^2/4}.
\end{equation}
After discretizing the real energy for scattering solutions and
phase shift variation, the resonance energy is uniquely
determined.

One can find from Tab.~\ref{tab1} that the results from the ACCC
and the scattering methods agree well. Different effective
interactions give similar energies and widths. Therefore in the
following calculations, we use only the effective interaction NL3
for illustration.

\begin{figure}[htb!]

   \centering

   {\includegraphics[height=7cm]{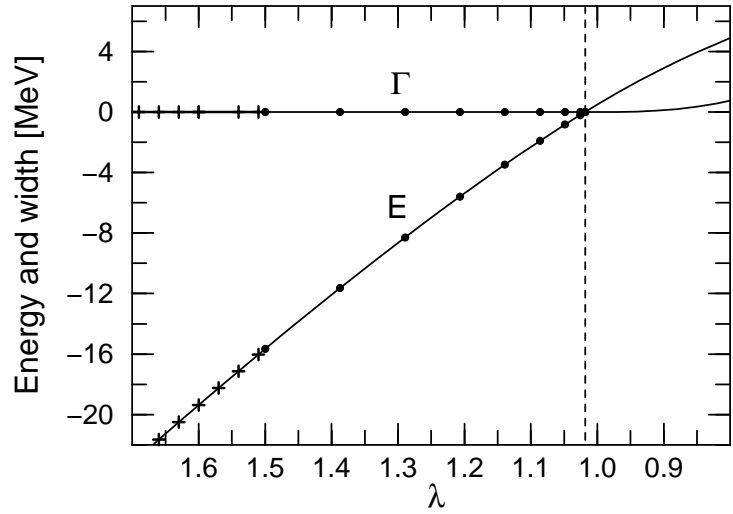} }\caption{ Energy $E$ and
   width $\Gamma$ for the neutron resonant state $\nu 1g_{9/2}$ in
   $^{60}$Ca, as a function of the coupling constant $\lambda$.
   Filled circles and crosses are the solutions of the Dirac
   equation, and the former is used as input in the ACCC method.
   Solid curves are the outcome of $L=N=4$ PA.} \label{fig-1}

\end{figure}

The resonance energy $E$ and width $\Gamma$ for the neutron
resonant state $\nu 1g_{9/2}$ in $^{60}$Ca, as a function of the
coupling constant $\lambda$, are exhibited in Fig.~\ref{fig-1}.
Filled circles and crosses are the solutions of the Dirac equation
[Eq. (\ref{Dirac-r})], and the former is used as input in the ACCC
method. Solid curves are the outcome of $L=N=4$ PA [Eq.
(\ref{pade-e})]. The dashed vertical line corresponds to the
branch point $\lambda_0 > 1$. When $\lambda > \lambda_0$, the
particle is bound. With $\lambda$ decreasing from $\lambda_0$ down
to 1, a nonvanishing imaginary part appears in the complex energy
expressed by Eq. (\ref{E-W}). At $\lambda = 1$, one gets the
resonance energy $E$ and width $\Gamma$. The agreement between the
exact solutions of the Dirac equation with large coupling constant
(crosses in Fig.~\ref{fig-1}) and the extrapolated results of PA
(solid energy curve) is quite good, which demonstrates the
validity of the ACCC method. The smooth solid line indicates good
analyticity of the eigenvalue for the Dirac equation with respect
to the coupling constant. The same results can also be obtained
for the case of $L=N=5$ and $L=N=6$ PAs, similar as in
Ref.~\cite{ZhangSch,ZhangDirac}.

\begin{figure}[htb!]

   \begin{minipage}{7.5cm}
       \centering {\includegraphics[height=10.0cm]{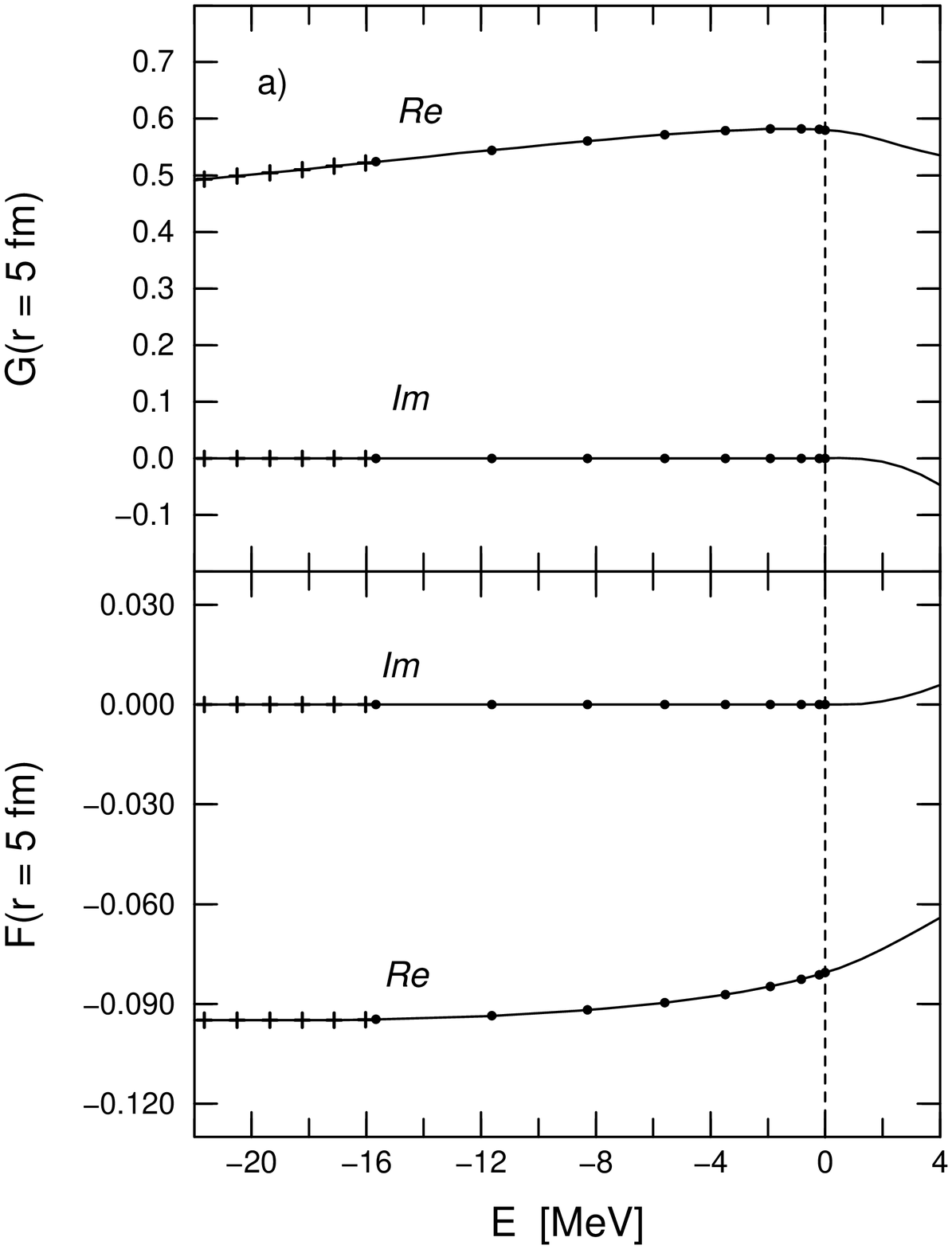} }
   \end{minipage}
   \hfill
   \begin{minipage}{7.5cm}
       \centering {\includegraphics[height=10.0cm]{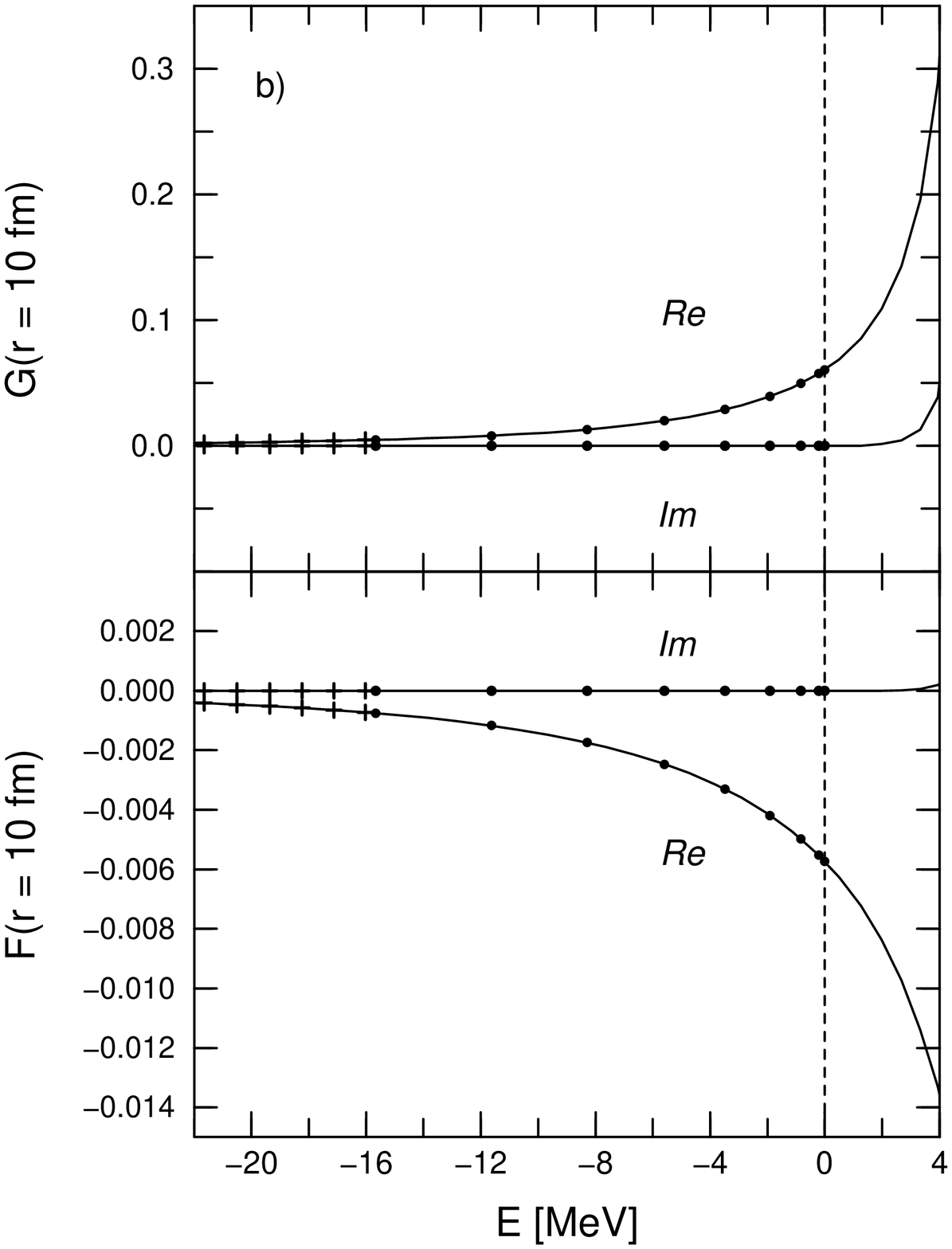} }
   \end{minipage}

   \caption{The real and imaginary part of the upper component $G(r)$
   and the lower component $F(r)$ for the neutron $\nu 1g_{9/2}$
   state in $^{60}$Ca at (a) $r=5$ fm and (b) $r=10$ fm as functions
   of the resonance energy $E$ (which corresponds to the coupling
   constant $\lambda$ displayed in Fig. 1). Symbols have the same
   meaning as those in Fig. 1. }

   \label{fig-2}

\end{figure}

In Fig.~\ref{fig-2}, we show the real and imaginary parts of the
upper component $G(r)$ and the lower component $F(r)$ at $r=5$ fm
and $r=10$ fm respectively as functions of the resonance energy
$E$ (which corresponds to the coupling constant $\lambda$
displayed in Fig. \ref{fig-1}). Symbols in Fig.~\ref{fig-2} have
the same meaning as those in Fig.~\ref{fig-1}. The neutron feels
much stronger effect from the nuclear potential at $r=5$ fm than
at $r=10$ fm. Similar to the case of the eigenvalue discussed
above, the real and the imaginary part of each component behave
smoothly, which indicates good analyticity of the eigenfunction
for the Dirac equation with respect to the coupling constant.

\begin{figure}[htb!]

    \centering
   {\includegraphics[height=12cm]{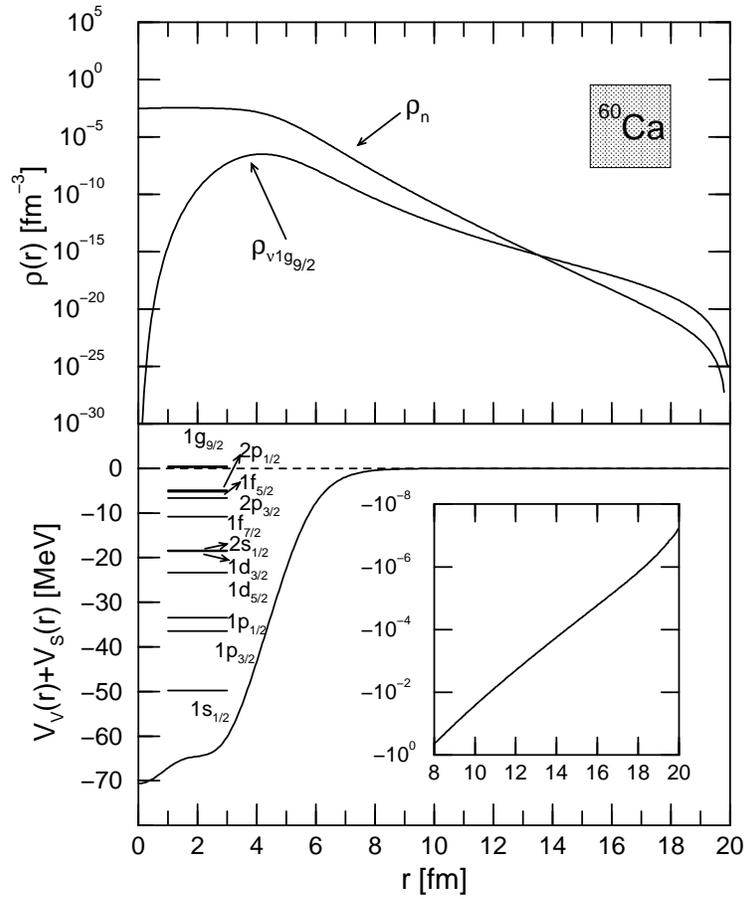} }\caption{Neutron
   densities for $^{60}$Ca and for the neutron $\nu 1g_{9/2}$ state
   (upper panel); the potential $V_V(r)+V_S(r)$, the single-particle
   neutron bound states and the neutron resonant state $\nu 1g_{9/2}$
   (lower panel).}

   \label{fig-3}

\end{figure}

\begin{figure}[htb!]

   \centering
  {\includegraphics[height=9cm]{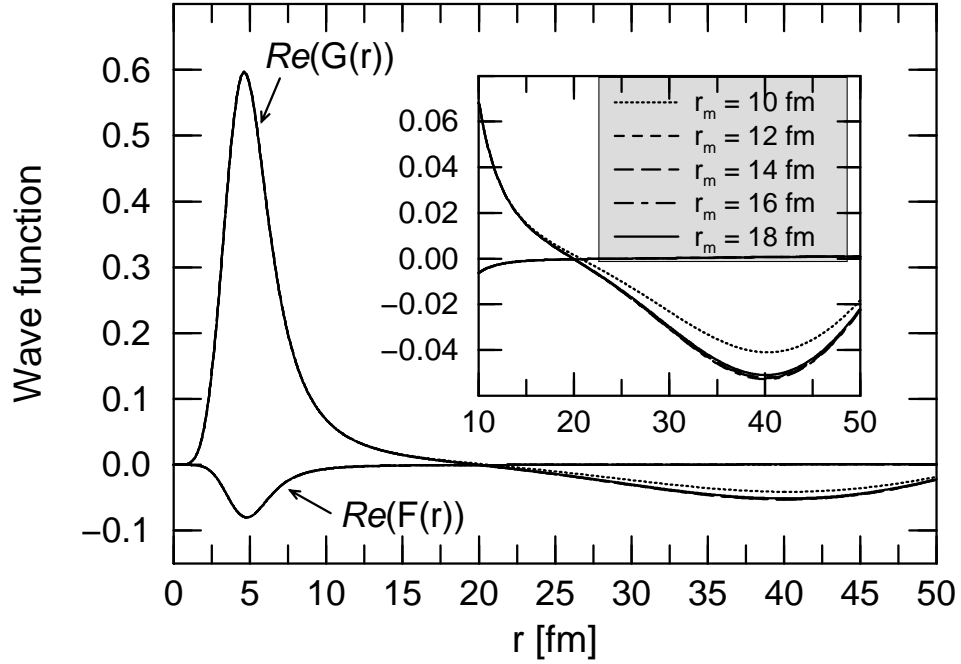} }\caption{ Real parts of
  the upper and lower components of the radial wave function for the
  neutron $\nu 1g_{9/2}$ state in $^{60}$Ca calculated by the ACCC
  method with different matching points $r_m$. }

  \label{fig-4}

\end{figure}

The wave function obtained from the analytic continuation is
connected continuously to the free wave function at the matching
point $r_m$. On one hand, the matching point $r_m$ should be large
enough to make sure that the potential at $r_m$ vanishes, which is
required by the outer part of the wave function. On the other
hand, $r_m$ should be far enough from the box radius $R_0$ to
guarantee that the inner part of the wave function can be obtained
accurately. In the lower panel of Fig.~\ref{fig-3}, the potential
$V_V(r)+V_S(r)$ for $^{60}$Ca is plotted. One finds that when $r >
14$~fm, the potential becomes smaller than $10^{-4}$ MeV which is
good enough for the precision in our calculations. The upper limit
of the matching point $r_m$ can be estimated by the behavior of
the density from the upper panel in Fig.~\ref{fig-3}. The neutron
density $\rho_{n}(r)$ in $^{60}$Ca and $\rho_{\nu 1g_{9/2}}(r)
\equiv [|G(r)|^2+|F(r)|^2]/4\pi r^2$ decline rapidly in the
vicinity of 20 fm because of the boundary condition in the RMF
calculation. The density curves in the range $r < 18$~fm are
smooth and behave reasonably. Therefore, the matching point should
be chosen at any point between 14~fm and 18~fm. We present the
wave function for the neutron resonant state $\nu 1g_{9/2}$ in
$^{60}$Ca with different matching points in Fig.~\ref{fig-4}. As
expected, the wave function is the same when changing the matching
point $r_m$ from 14 fm to 18 fm. In the following calculations,
the wave function will be matched at $r_m = 18$~fm, unless
otherwise specified.

\section{Application to single-particle resonant states in
$^{122}$Zr}

We apply the ACCC method within the framework of the RMF theory to
calculate the energies, widths and wave functions for the resonant
states in $^{122}$Zr. Since $^{122}$Zr is the core of the giant
neutron halo predicted in RCHB calculations~\cite{MR98}, and the
resonant states in $^{122}$Zr have been obtained by the scattering
method~\cite{San}, we choose this nucleus for comparison.

\begin{figure}[htb!]

    \centering
   {\includegraphics[height=9cm]{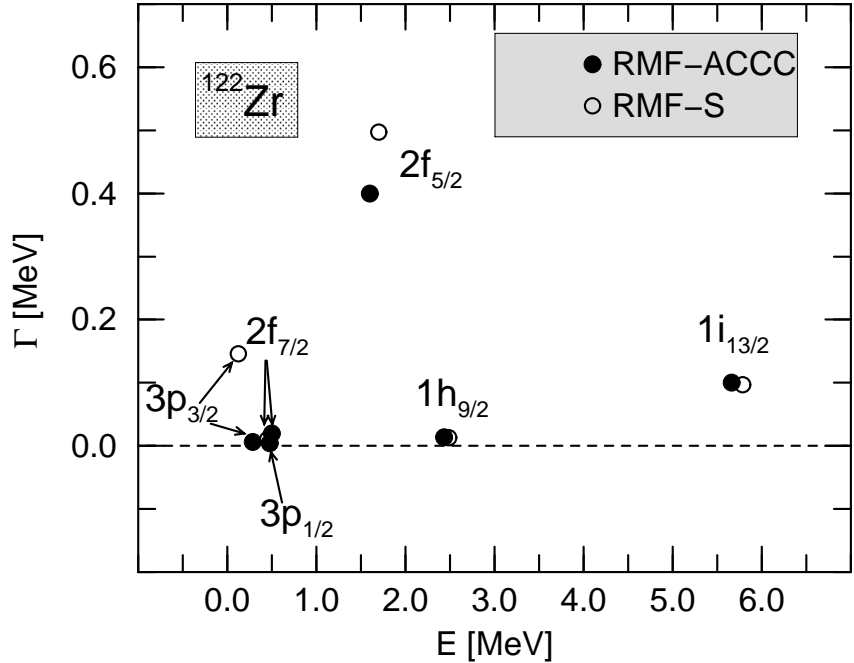} }\caption{ Energies and
   widths for the neutron state $\nu 3p_{3/2}$, $\nu 3p_{1/2}$, $\nu 2f_{7/2}$,
   $\nu 2f_{5/2}$, $\nu 1h_{9/2}$, and $\nu 1i_{13/2}$ in $^{122}$Zr. Solid
   circles represent the results of the ACCC method, while open
   circles denote the results of the scattering method.}

   \label{fig-5}

\end{figure}

In Fig.~\ref{fig-5}, the energies and widths for the resonant
states $\nu 3p_{3/2}$, $\nu 3p_{1/2}$, $\nu 2f_{7/2}$, $\nu
2f_{5/2}$, $\nu 1h_{9/2}$ and $\nu 1i_{13/2}$ are presented as a
planar $E$--$\Gamma$ plot. The results of the ACCC and the
scattering methods are in good agreement with each other for most
of these states. From both methods, $\nu 2f_{5/2}$ and $\nu
1i_{13/2}$ have large widths, while $\nu 2f_{7/2}$ and $\nu
1h_{9/2}$ are very narrow. The width for $\nu 3p_{3/2}$ from the
scattering method is slightly larger than our calculation. As for
the resonant state $\nu 3p_{1/2}$, neither the energy nor the
width can be extracted from the scattering calculation. From a
quantum mechanical point of view, these single-particle resonant
states are quasi-stationary ones captured by centrifugal barriers.
The decay width for a resonant state can be roughly explained by
the penetration through the barrier. For those states with the
same $l$, i.e., the same centrifugal barrier, the higher state has
a larger width, e.g., for the two $f$ states $\nu 2f_{5/2}$ and
$\nu 2f_{7/2}$. Although $\nu 1h_{9/2}$ is higher than $\nu
2f_{5/2}$, its width is smaller because of higher centrifugal
barrier. A similar argument also holds for a much higher but
narrow state $\nu 1i_{13/2}$.

\begin{figure}[htb!]
    \centering
   {\includegraphics[height=9cm]{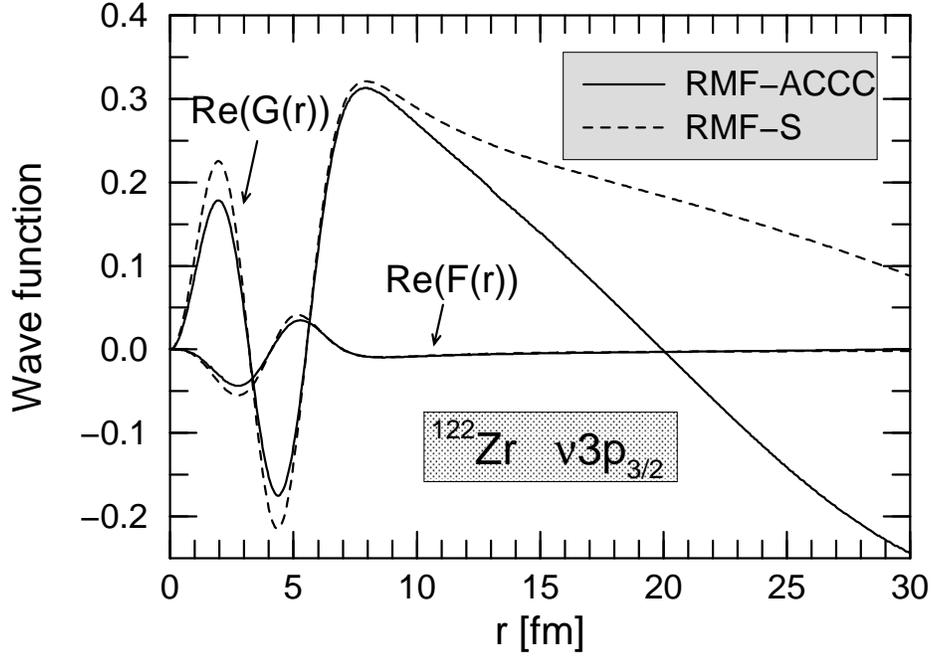} }\caption{ Real parts of
   the upper and lower components of radial wave functions for the
   neutron resonant state $\nu 3p_{3/2}$ in $^{122}$Zr. Solid and dashed
   curves represent the results of the ACCC and the scattering
   method respectively. }

   \label{fig-6}

\end{figure}

\begin{figure}[htb!]

    \centering
   {\includegraphics[height=9cm]{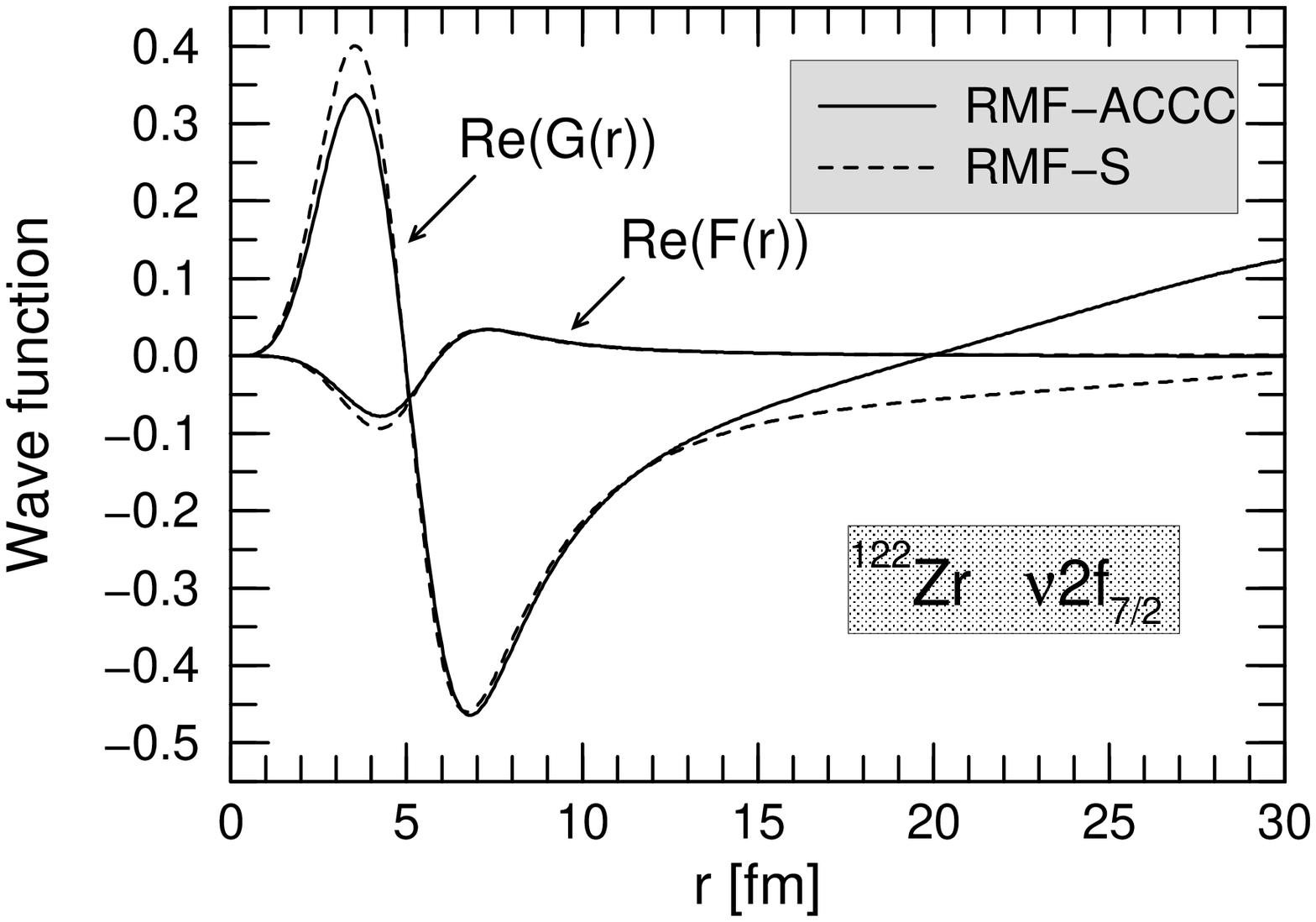} }\caption{ Similar as Fig.
   6, but for the neutron resonant state $\nu 2f_{7/2}$ in $^{122}$Zr.}

   \label{fig-7}

\end{figure}

\begin{figure}[htb!]

     \centering
    {\includegraphics[height=9cm]{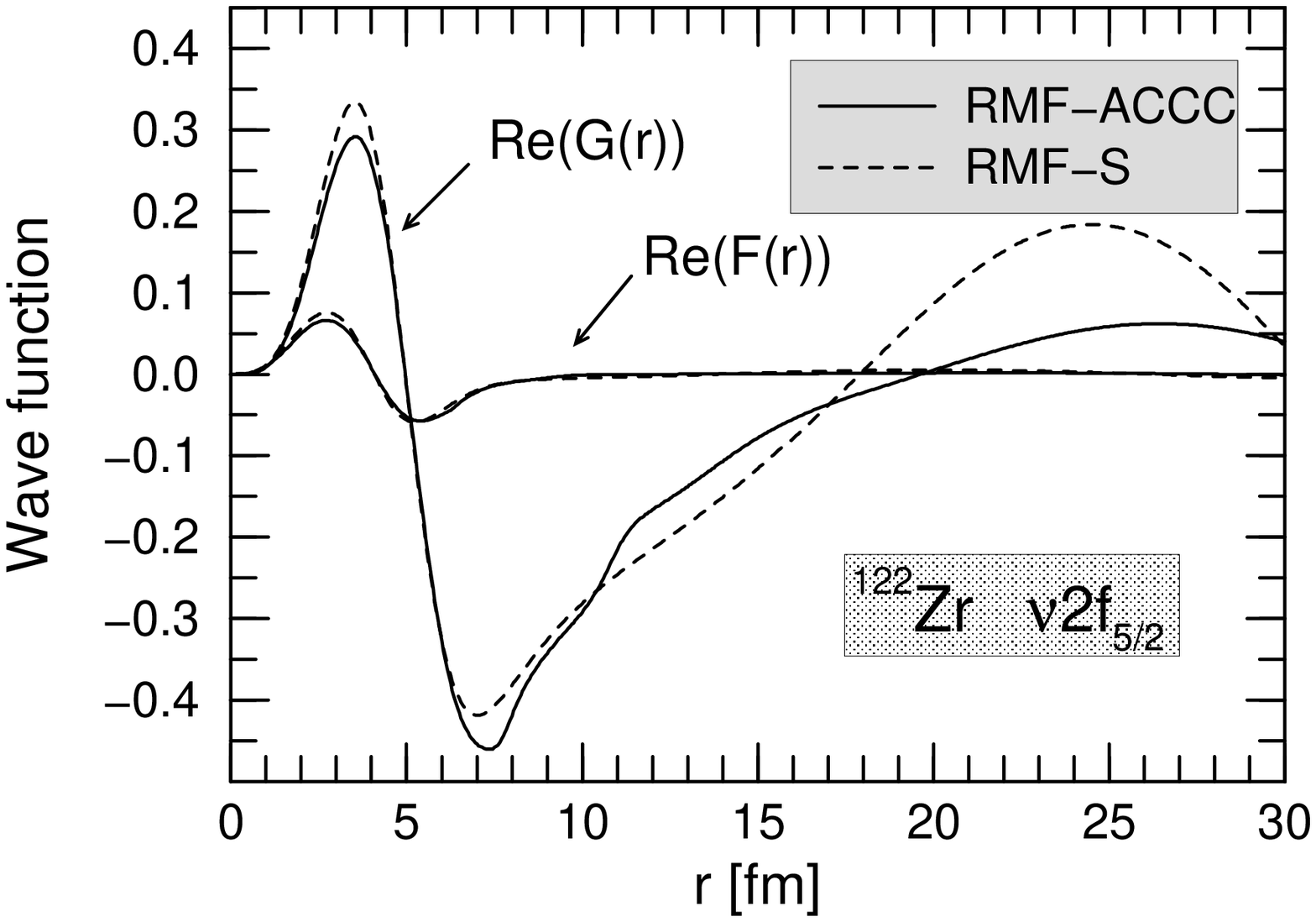} }\caption{Similar as Fig.
    6, but for the neutron resonant state $\nu 2f_{5/2}$ in $^{122}$Zr.}

   \label{fig-8}

\end{figure}

\begin{figure}[htb!]
    \centering
    {\includegraphics[height=9cm]{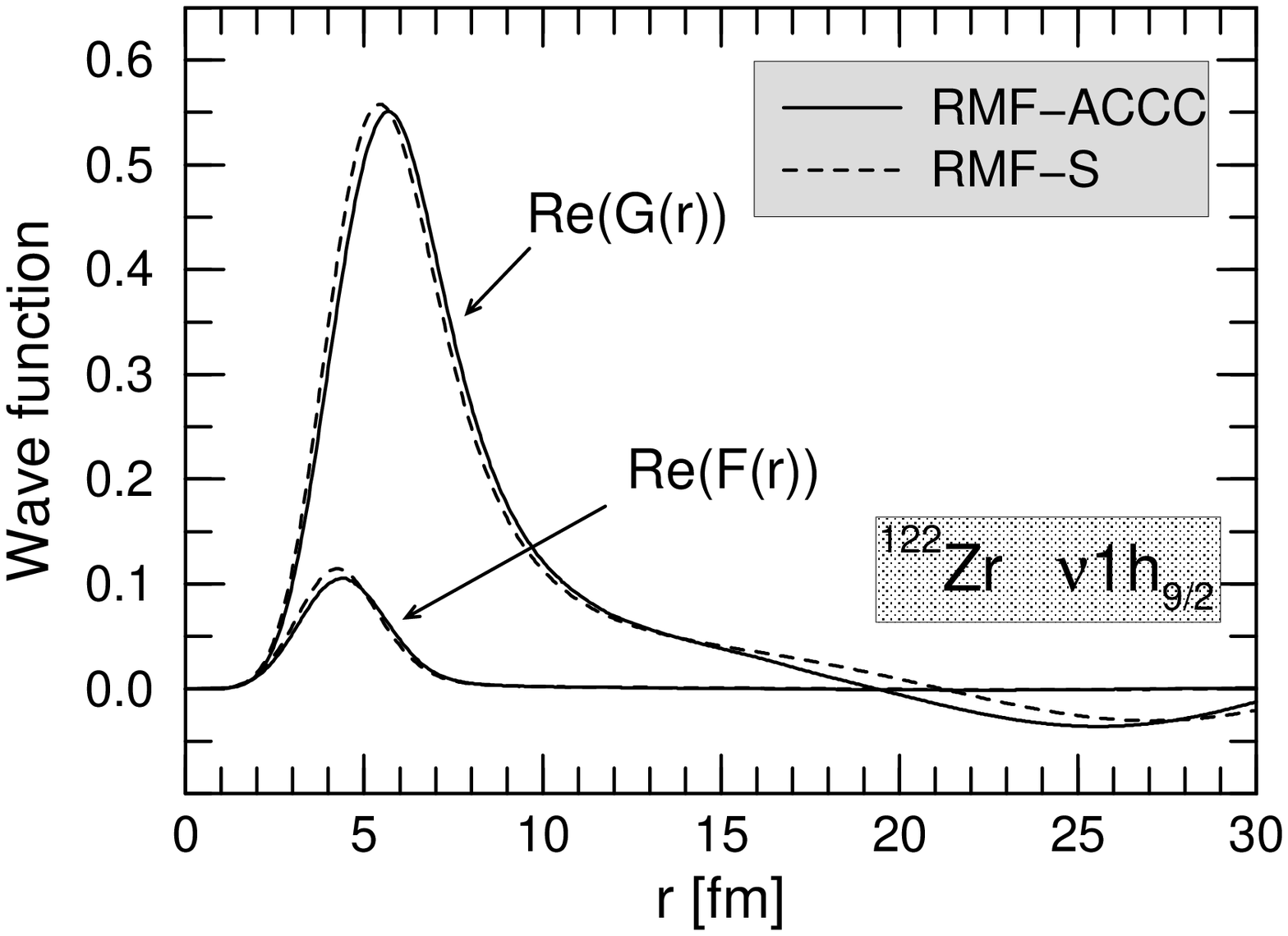} }\caption{Similar as  Fig.
    6, but for the neutron resonant state $\nu 1h_{9/2}$ in $^{122}$Zr.}

    \label{fig-9}

\end{figure}

The wave functions for the resonant states $\nu 3p_{3/2}$, $\nu
2f_{7/2}$, $\nu 2f_{5/2}$ and $\nu 1h_{9/2}$ are respectively
exhibited in Figs.~\ref{fig-6}, \ref{fig-7}, \ref{fig-8}, and
\ref{fig-9}. Since the imaginary part of the wave function is
neglected in the scattering calculations~\cite{San,San00}, only
the real parts of the upper and the lower components are given and
compared in these figures. In Fig.~\ref{fig-6}, one finds a
considerable difference for the upper components at large $r$ in
$\nu 3p_{3/2}$ between these two methods. This difference is
consistent with the difference in energy and width as seen in
Fig.~\ref{fig-5}. A similar difference is also seen for $\nu
2f_{5/2}$ in Fig.~\ref{fig-8}. For those states where the two
methods give nearly the same energies and widths, e.g., $\nu
2f_{7/2}$ and $\nu 1h_{9/2}$, the wave functions agree quite well.
The small difference for the wave functions between these two
methods found in Figs.~\ref{fig-7} and \ref{fig-9} might be from
the neglect of the imaginary part of the wave function in the
scattering method.

\section{Summary}

In summary, the method of the analytic continuation in the
coupling constant (ACCC) has been developed within the framework
of the relativistic mean field (RMF) theory and used to
investigate the resonance energies, widths and wave functions for
single-particle resonant states in exotic nuclei. The analyticity
of these quantities, as functions of the coupling constant, are
exploited for the neutron resonant state $\nu 1g_{9/2}$ in
$^{60}$Ca. The energies, widths and wave functions for some
resonant states in $^{122}$Zr are calculated and compared with
those obtained from the scattering method. The results from the
ACCC method agrees satisfactorily with those from the scattering
calculation as well as available data ~\cite{Yang01}. It is
expected that the RMF-ACCC method, combined with the BCS theory,
can be generalized to describe exotic phenomena in unstable
nuclei.

\begin{acknowledgments}

This work is partly supported by the Major State Basic Research
Development Program Under Contract Number G2000077407, the
National Natural Science Foundation of China under Grant No.
10025522, and 10221003, and the Doctoral Program Foundation from
the Ministry of Education in China under Grant No. 20030001086, as
well as the National Science Foundation of South Africa under
Grant No. 2054166.

\end{acknowledgments}

\newpage

\begin{table}[h]

    \vspace{0.5cm} \centering \caption{Energies and widths
   ($E,~\Gamma)$ in MeV for the neutron resonant state $\nu 1g_{9/2}$
   in $^{60}$Ca calculated by the ACCC (RMF-ACCC) and the scattering
   (RMF-S) method with the effective interactions NL3, NLSH, PK1, and
   TMA.} \vspace{0.5cm}

   \begin{tabular}{c|@{\hspace{0.8cm}}c@{\hspace{0.8cm}}c@{\hspace{0.8cm}}c@{\hspace{0.8cm}}c}\toprule\hline

      &NL3&NLSH&PK1&TMA\\

      \midrule\hline

      RMF-ACCC [MeV] &(0.49, 0.0005)& (0.47, 0.0004) &(0.77, 0.0009) & (0.77, 0.0009) \\

      RMF-S [MeV] &(0.54, 0.0001)& (0.53, 0.0001) &(0.85, 0.0006)&(0.83, 0.0005)\\

      \bottomrule\hline

   \end{tabular}

   \label{tab1}

\end{table}

\begin{thebibliography}{99}
\bibliographystyle{unsrt}

\bibitem{TA85} I. Tanihata, H. Hamagaki, and O. Hashimoto et al. Phys. Rev. Lett. {\bf 55}, 2676 (1985).

\bibitem{MR96} J. Meng and P. Ring, Phys. Rev. Lett. {\bf 77}, 3963 (1996).

\bibitem{MR98} J. Meng and P. Ring, Phys. Rev. Lett. {\bf 80}, 460 (1998).
               J. Meng, H. Toki, J. Y. Zeng, S. Q. Zhang, and S. G. Zhou, Phy. Rev. C {\bf 65}, 041302(R) (2002).

\bibitem{ME98} J. Meng, Nucl. Phys. {\bf A635}, 31 (1998).

\bibitem{San00} N. Sandulescu, Nguyen Van Giai, and R. J. Liotta, Phy. Rev. C {\bf 61}, 061301(R) (2000).

\bibitem{San} N. Sandulescu, L. S. Geng, H. Toki, and G. Hillhouse, Phys. Rev. C {\bf 68}, 054323 (2003).

\bibitem{Curutchet89} P. Curutchet, T. Vertse, and R. J. Liotta, Phys. Rev. C {\bf 39}, 1020 (1989).

\bibitem{Cao02} L. G. Cao and Z. Y. Ma, Phys. Rev. C {\bf 66}, 024311 (2002).

\bibitem{Wigner47} E. Wigner and L. Eisenbud, Phys. Rev. {\bf 72}, 29 (1947).

\bibitem{Hale87} G. M. Hale, R. E. Brown, and N. Jarmie, Phys. Rev. Lett. {\bf 59}, 763 (1987).

\bibitem{Humblet91} J. Humblet, B. W. Filippone, and S. E. Koonin, Phys. Rev. C {\bf 44}, 2530 (1991).

\bibitem{Taylor72} J. R. Taylor, {\it Scattering Theory: The Quantum Theory on Nonrelativistic Collisions},
(John Wiley \& Sons, New York, 1972).

\bibitem{Dobaczewski84} J. Dobaczewski, H. Flocard, and J. Treiner, Nucl. Phys. {\bf A422}, 103 (1984).

\bibitem{Hazi70} A. U. Hazi and H. S. Taylor, Phys. Rev. A {\bf 1}, 1109 (1970).

\bibitem{Ho83} Y. K. Ho, Phys. Rep. {\bf 99}, 1 (1983).

\bibitem{kukulin89} V. I. Kukulin, V. M. Krasnopl'sky and J. Hor$\acute{a}$cek,
{\it Theory of Resonances: Principles and Applications} (Kluwer
Academic, Dordrecht, 1989).

\bibitem{tanaka99} N. Tanaka, Y. Suzuki, and K. Varga et al. Phys. Rev. C {\bf 59}, 1391 (1999).

\bibitem{Aoyama02} S. Aoyama, Phys. Rev. Lett. {\bf 89}, 052501 (2002).

\bibitem{cattapan} G. Cattapan and E. Maglione, Phys. Rev. C {\bf 61}, 067301 (2000).

\bibitem{ZhangSch} S. S. Zhang, J. Meng, and J. Y. Guo, High Ener. Phys. and Nucl. Phys. {\bf 27}, 1095 (2003) (in Chinese).

\bibitem{ZhangDirac} S. S. Zhang, J. Y. Guo, S. Q. Zhang, and J. Meng, Chin. Phys. Lett. {\bf 21}, 632 (2004).

\bibitem{Ring96} P. Ring, Prog. Part. Nucl. Phys. {\bf 37}, 193 (1996).

\bibitem{Yang01} S. C. Yang, J. Meng, and S. G. Zhou, Chin. Phys. Lett. {\bf 18}, 196 (2001).

 \bibitem{BB77} J. Boguta and A. R. Bodmer, Nucl. Phys. {\bf A292} (1977) 413.

\bibitem{ME99} J. Meng, K. Sugawara-Tanabe, S. Yamaji, P. Ring, and A. Arima, Phys. Rev. C {\bf 58}, 628(R) (1998).
 J. Meng, K. Sugawara-Tanabe, S. Yamaji, and A. Arima, Phys. Rev. C {\bf 59}, 154 (1999).

\bibitem{ZMR03} S. G. Zhou, J. Meng, and P. Ring, Phys. Rev. C {\bf 68}, 034323 (2003).

\bibitem{NL3}G. A. Lalazissis, J. K\"onig, and P. Ring, Phys. Rev. C {\bf 55}, 540 (1997).

\bibitem{NLSH} M. M. Sharma, M. A. Nagarajan, and P. Ring, Phys. Lett. B {\bf 312}, 377 (1993)

\bibitem{TMA} Y. Sugahara and H. Toki, Nucl. Phys. {\bf A579}, 557 (1994).

\bibitem{PK1} W. H. Long, J. Meng, N. V. Giai, and S. G. Zhou. Phys. Rev. C (in press). [arXiv: nucl-th/0311031]

\end{thebibliography}
\end{document}